\documentclass[aps,prd,twocolumn,groupedaddress]{revtex4-2}

\usepackage{amssymb}
\usepackage{amsmath}
\usepackage{graphicx}
\usepackage{enumerate}
\usepackage{mathtools}
\usepackage{color}
\usepackage{bbold}
\usepackage{subfigure}
\usepackage{slashed}
\usepackage{xcolor}
\usepackage{bm}
\usepackage{comment}

\begin{document}


\title{Fast flavor instabilities and the search for neutrino angular crossings}


\author{Lucas Johns}
\email[]{NASA Einstein Fellow (ljohns@berkeley.edu)}
\affiliation{Departments of Astronomy and Physics, University of California, Berkeley, CA 94720, USA}

\author{Hiroki Nagakura}
\affiliation{Department of Astrophysical Sciences, Princeton University, Princeton, NJ 08544, USA}

\begin{abstract}
With the recognition that fast flavor instabilities likely affect supernova and neutron-star-merger neutrinos, using simulation data to pin down when and where the instabilities occur has become a high priority. The effort faces an interesting problem. Fast instabilities are related to neutrino angular crossings, but simulations often employ moment methods, sacrificing momentum-space angular resolution in order to allocate resources elsewhere. How can limited angular information be used most productively? The main aims here are to sharpen this question and examine some of the available answers. A recently proposed method of searching for angular crossings is scrutinized, the limitations of moment closures are highlighted, and two ways of reconstructing angular distributions solely from the flux factors (based respectively on maximum-entropy and sharp-decoupling assumptions) are compared. In (semi)transparent regions, the standard closure prescriptions likely miss some crossings that should be there and introduce others that should not.
\end{abstract}

\maketitle

\section{Introduction \label{sec:intro}}

Calculations predict that supernova and neutron-star-merger neutrinos experience fast flavor instabilities at some times and locations, possibly with important consequences for the dynamics, composition, and neutrino signal \cite{sawyer2005, sawyer2009, sawyer2016, chakraborty2016b, chakraborty2016c, dasgupta2017, tamborra2017, wu2017, wu2017b, dasgupta2018, dasgupta2018b, airen2018, abbar2018, abbar2019, abbar2019b, azari2019, shalgar2019, nagakura2019, capozzi2019, capozzi2019b, yi2019, chakraborty2020, martin2020, johns2020, johns2020b, shalgar2020b, capozzi2020, xiong2020, shalgar2020, shalgar2020c, bhattacharyya2020b, morinaga2020, glas2020, abbar2020b, padilla2020, george2020, martin2021, bhattacharyya2021, bhattacharyya2021b, richers2021, tamborra2020, abbar2020, abbar2021, capozzi2021, li2021, morinaga2021}. Of the physical conditions that affect stability, the electron-lepton-number (ELN) angular distribution $G(\mu)$ is recognized as being particularly important:
\begin{equation}
G(\mu) =  \int_0^{\infty} \frac{dE_\nu E_\nu^2}{(2\pi)^2} \left( g_{\nu_e} (E_\nu, \mu) - g_{\bar{\nu}_e} (E_\nu, \mu) \right),
\end{equation}
where $g_{\nu_e}$ and $g_{\bar{\nu}_e}$ are the neutrino and antineutrino distribution functions, $E_\nu$ is neutrino energy, and $\mu = \cos\theta$ is the angle of propagation away from the radial direction. (Axial symmetry is assumed and differences among the heavy-lepton flavors are ignored.) Angular crossings of $G(\mu)$---points where the function passes through 0---are associated with fast instability.

Since the presence of angular crossings is a simple criterion, searching for crossings is a useful alternative to running every single location through linear stability analysis. But while the criterion is straightforward to apply if the neutrino angular distributions are known, they usually are not. Because of the high complexity of supernova and merger physics, sacrifices simply have to be made in some aspects of the problem, even when using the most powerful supercomputers. Computing neutrino transport with high angular resolution in momentum space---that is, finely resolving the $\mathbf{\hat{p}}$-dependence of the neutrino distributions---is expensive, and many radiation-hydrodynamics simulations opt for evolving only the lowest moments of the angular distributions (the energy densities and fluxes, for example). Thus we face the question that occupies us here: \textit{How can this limited amount of information best be used in the search for angular crossings?}

Recently a few publications have advocated and adopted a search procedure that involves applying polynomial weighting functions to $G(\mu)$ \cite{abbar2020, abbar2021, capozzi2021}. In Sec.~\ref{sec:points} we examine this approach. We find that it never introduces spurious crossings but sometimes misses physical ones. In our view, more evidence is needed to establish whether this approach has any advantages over searching point-by-point for crossings in $G(\mu)$.

In Sec.~\ref{sec:criteria}, inspired by Ref.~\cite{abbar2020}'s contrasting of the polynomial method with other criteria found in the literature, we offer our own remarks on the similarities and differences.

Regardless of how the search is conducted, it inevitably runs into problems in (semi)transparent regions if only the first few angular moments are employed. In Sec.~\ref{sec:closures} we demonstrate this point, emphasizing that a closure prescription can be adequate for energy--momentum transport while being inadequate for angular-crossing searches. The paradigmatic example is a highly forward-peaked distribution. This case is easy to describe from the transport perspective: particles travel like a beam, and pressure is nonzero only along the flux direction. But the same distribution flusters angular moments, as very many of them are required to specify an approximation that is not compromised by unphysical features like negative particle numbers.

In this paper we use \textit{completion} to mean a prescription for reconstructing angular distributions from the first few moments (or, equivalently, generating all higher moments). We use this term in contrast with \textit{closure}, by which we mean a prescription for generating a finite number of higher moments. In practice, closures usually only specify moments that are one or two ranks above those that are actually being evolved, because that suffices to close the hierarchy of coupled transport equations. Our goal, however, is not to approximate transport by truncating the hierarchy of equations: our goal is to fill in the details of the angular distributions. We would prefer to complete, rather than close, the angular distributions that are output by moment-based simulations.

Here we focus on two completions that depend only on the neutrino flux factor $f = F / E$, where $F$ is the energy flux and $E$ is the energy density. The \textit{maximum-entropy completion} fills in the details by assuming that the angular distribution is of an entropy-maximizing form. The same distribution types have been used to motivate closures; here we are simply adopting them for a different (but obviously related) purpose. We compare this first prescription with the \textit{sharp-decoupling completion}, which maps an angular distribution with flux factor $f$ onto the angular distribution in the supernova bulb model with the same flux factor. These two completions are presented in Sec.~\ref{sec:completions}.

We also find in Sec.~\ref{sec:completions} that, given the same 0th and 1st moments, closed and completed distributions do not always agree on whether a crossing occurs. The disagreement can go either way, with closed distributions exhibiting a crossing not found with completed distributions or vice versa. In view of this fact, we urge caution when presenting or interpreting stability analyses that apply closures to obtain the angular distributions.

The maximum-entropy and sharp-decoupling completions have their own strengths and weaknesses, and we are not endorsing either one as a final answer to the question of how best to search for angular crossings using moment data. In Sec.~\ref{sec:discussion} we indicate the steps we will take elsewhere to provide a more actionable answer to this question.

\section{Points vs. polynomials \label{sec:points}}

In the method proposed in Ref.~\cite{abbar2020} and utilized in Refs.~\cite{abbar2021, capozzi2021}, a weighting function is introduced,
\begin{equation}
\mathcal{F}(\mu) = \sum_{n=0}^N a_n \mu^n \label{Fmudef},
\end{equation}
such that $\mathcal{F}(\mu) > 0$ for all $\mu \in [-1,1]$. The coefficients $a_n$ are arbitrary aside from having to satisfy the positivity constraint, and $N$ labels the highest angular moment for which simulation data is available. A weighted version of the ELN angular distribution is then formed,
\begin{equation}
I_\mathcal{F} = \int_{-1}^{1} d\mu ~\mathcal{F}(\mu) G(\mu), \label{IFdef}
\end{equation}
and the search for angular crossings is conducted by searching for functions $\mathcal{F}(\mu)$ that satisfy
\begin{equation}
I_\mathcal{F} I_0 < 0, \label{IFI0}
\end{equation}
where
\begin{equation}
I_n = \int_{-1}^{1} d\mu ~ \mu^n G(\mu).
\end{equation}
The idea is that if there is an angular crossing, a well-chosen polynomial $\mathcal{F}(\mu)$ will exaggerate the region where the crossing occurs, causing the integrated quantity $I_\mathcal{F}$ to have a different sign than $I_0$. We call this search procedure the \textit{polynomial method}.

An alternative is to simply scan over $G(\mu)$, evaluating the function at different values of $\mu$ and checking if it changes sign (the \textit{point-by-point method}). Ref.~\cite{abbar2020} does not explain why (or whether) the polynomial method is superior to the point-by-point method---the latter procedure is not mentioned---but the paper does observe, just as $\mathcal{F}(\mu)$ is being introduced (footnote 2), that an angular-moment expansion cannot be used to approximate $G(\mu)$ because higher moments are not necessarily negligible---they simply are not provided by moment-based simulations. We agree with this point and develop it further in Sec.~\ref{sec:closures}.

However, the polynomial method does not evade the issue of non-negligible higher moments. The product $\mathcal{F}(\mu) G(\mu)$ that appears in Eq.~\eqref{IFdef} is a kind of reshaped version of $G(\mu)$, and so it might be thought that the method in some way explores a fuller range of ELN angular distributions (because of the flexibility in $\mathcal{F}$) while at the same time being informed by the simulation's output (because the starting point is $G$). This is not the case, unfortunately. Let $G'(\mu) \equiv \mathcal{F}(\mu) G(\mu)$ be the reshaped ELN angular distribution. Then
\begin{equation}
I_\mathcal{F} = \int_{-1}^1 d\mu~G'(\mu) = I'_0,
\end{equation}
and the criterion $I_\mathcal{F} I_0 < 0$ becomes $I'_0 I_0 < 0$, which simply checks if $G(\mu)$ is capable of being reshaped in such a way that the overall lepton number changes sign. The polynomial method is thus in no way akin to searching for crossings in physically motivated modifications of the moment-reconstructed distribution. In fact the procedure is in quite the opposite spirit: the distribution is reshaped so as to \textit{artificially} exaggerate crossing regions. The polynomial and point-by-point methods are equally limited by truncation of the angular-moment expansion.

Moreover, the polynomial method sometimes misses angular crossings. (Because $\mathcal{F}(\mu) > 0$, spurious crossings are never created by the search procedure itself, although they may appear due to neglecting higher moments in the ELN angular distribution.) As an example of missed crossings, let $N = 1$, define $\epsilon$ by
\begin{equation}
\frac{I_1}{I_0} = \frac{1}{3} + \epsilon, \label{epsdef}
\end{equation}
and consider the case with $I_0 > 0$. Since
\begin{equation}
G(\mu) = \frac{I_0}{2} \left( 1 + 3\mu \frac{I_1}{I_0} \right),
\end{equation}
a crossing exists in the backward direction for $\epsilon > 0$. The criterion $I_\mathcal{F} I_0 < 0$ can be rewritten as $I_\mathcal{F} / I_0 < 0$, giving
\begin{equation}
a_1 < - \frac{a_0}{\frac{1}{3} + \epsilon} ~ \left[ \textrm{for crossing to be found} \right]. \label{N2crit}
\end{equation}
Assume that $a_0 > 0$. Then Eq.~\eqref{N2crit} entails
\begin{equation}
\mathcal{F}(\mu = +1) < a_0 \left( 1 - \frac{1}{\frac{1}{3} + \epsilon} \right).
\end{equation}
This says that if $0 < \epsilon < 2/3$, then $\mathcal{F}(\mu = +1)$ must drop below 0 in order to satisfy $I_\mathcal{F} I_0 < 0$. If instead $a_0 < 0$, then $\mathcal{F}(\mu = -1) < 0$ for all $\epsilon > 0$. Therefore many ELN angular distributions with crossings will not be identified as such in the polynomial search. The requisite polynomials are excluded by the positivity condition.

\begin{figure}
\centering
\includegraphics[width=.43\textwidth]{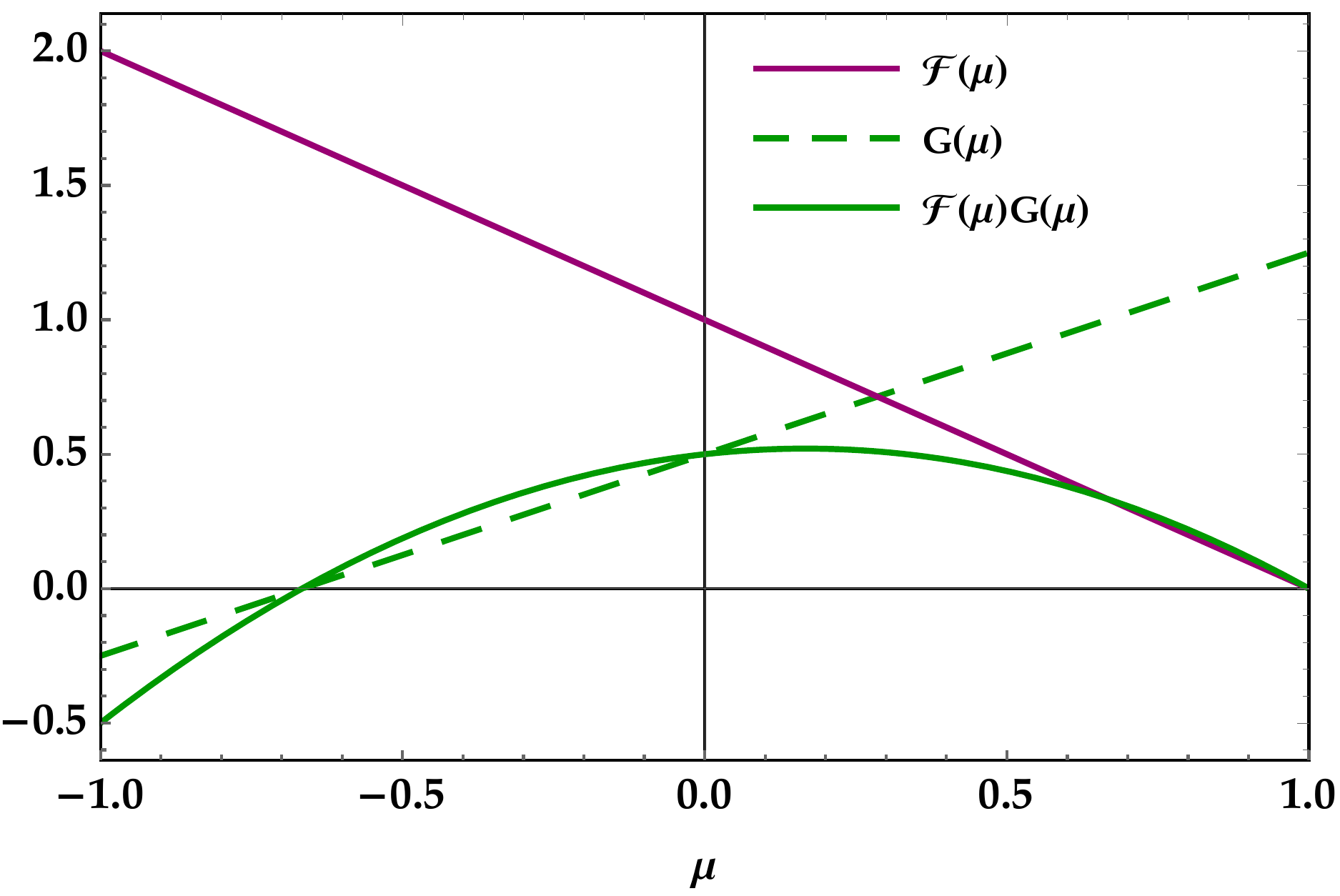}
\caption{$G(\mu)$ (dashed green curve), $\mathcal{F}(\mu)$ (purple), and $\mathcal{F}(\mu)G(\mu)$ (solid green) vs. $\mu = \cos\theta$, adopting $\epsilon = 1/6$ (Eq.~\eqref{epsdef}). The angular crossing at $\mu \cong -0.7$ is not detected by the polynomial search.}
\label{missed_crossing}
\end{figure}

The idea being demonstrated here is that because $\mathcal{F}(\mu)$ is required to be nonnegative, it is limited in how much it can amplify certain regions relative to other ones. The ELN angular distribution itself is not limited in this way, and so small enough crossings can go unnoticed. This intuitive point is illustrated in Fig.~\ref{missed_crossing}. $\mathcal{F} (\mu)$, the purple curve, weights the backward direction as much as possible relative to the forward direction, hoping to amplify the crossing region and make $I_\mathcal{F}$ go negative, but $\mathcal{F}(\mu) G(\mu)$ can only be distorted so much. The area under the solid green curve remains positive.

Based on the considerations above, we are unable to see any advantages to the polynomial method as compared to scanning point-by-point over $G(\mu)$. The latter is in fact equivalent to using delta functions in place of $\mathcal{F}(\mu)$, and because the philosophy of the polynomial method is that crossing regions should be amplified as much as possible, delta functions would intuitively seem to be the best way to accomplish this. The point-by-point method could also very well be more efficient, since the scan is directly over $\mu$ rather than over $N$th-degree polynomials with coefficients satisfying a constraint that must itself be satisfied at all $\mu \in [-1,1]$.

The possibility certainly remains that we have overlooked something in our assessment. Since the polynomial method has now been used in multiple publications on fast instabilities, it would be a valuable contribution to the literature for any advantages we may have overlooked to be spelled out.

\section{Zero modes, pendulums, \& polynomials \label{sec:criteria}}

Seeing as how several other instability criteria can be found in the literature, it may be worthwhile to clarify how they all relate to one another.

The most rigorous test of stability involves linearizing the equations of motion---more particularly, making them linear in the off-diagonal elements of the density matrices---and searching for normal-mode solutions of the form $e^{- i \Omega t + i K x}$, where for simplicity we assume a single spatial dimension \cite{banerjee2011,izaguirre2017}. The result is a dispersion relation that involves integrals such as
\begin{equation}
\int_{-1}^1 d\mu \frac{\mu^m}{w + z \mu},
\end{equation}
where $w = w (\Omega)$ and $z = z (K)$. Generally the dispersion relation is therefore transcendental and must be solved numerically.

Ref.~\cite{dasgupta2018} pointed out that by picking $K$ such that $z = 0$, the dispersion relation simplifies from a transcendental equation to a quadratic one, allowing one to solve for $\Omega (K_0)$ analytically. ($K_0$, the \textit{zero mode}, is the value of $K$ for which $z = 0$.) The criterion $\textrm{Im}~\Omega \neq 0$ can then be written as
\begin{equation}
\left( I_0 + I_2 \right)^2 - 4 I_1^2 < 0. \label{zero_mode}
\end{equation}

Ref.~\cite{johns2020} derived two different criteria that apply to homogeneous ($K = 0$) fast instabilities. The first of these comes from the demand that there be a trajectory $\tilde{v} \in [-1, 1]$ that satisfies a certain kind of resonance condition in the linear regime. If such a trajectory exists, the system is unstable. The criterion is simply
\begin{equation}
I_2^2 < I_1^2. \label{pend1}
\end{equation}
The other criterion comes from the pendulumlike nature of fast flavor conversion and reads
\begin{equation}
I_2^2 \leq \frac{4}{5} I_1 \left( 5 I_3 - 3 I_1 \right). \label{pend2}
\end{equation}
In Ref.~\cite{johns2020} these criteria served the purpose of demonstrating the validity of the paper's analysis, showing in particular how the analysis allows for criteria to be formulated that improve (in the homogeneous setting considered) upon the angular-crossing criterion. They can be viewed as being complementary to the zero-mode test of Ref.~\cite{dasgupta2018}, but one should be cautious in employing them on simulation data because, unlike the zero-mode test, they do not derive from exactly solving a particular mode's dispersion relation. One should also be cautious in applying the zero-mode test, of course, since it only assesses the stability of a single mode.

Ref.~\cite{abbar2020} claims that the polynomial search ``offers an infinite number of inequalities similar to'' Eq.~\eqref{zero_mode} and suggests that this particular inequality corresponds to $\mathcal{F}_\pm(\mu) = \mu^2 \pm 2 \mu + 1$. While it is true that
\begin{equation}
I_{\mathcal{F}_+} I_{\mathcal{F}_-} = \left( I_0 + I_2 \right)^2 - 4 I_1^2,
\end{equation}
this expression does not generally show up in the search because
\begin{equation}
I_\mathcal{F} I_0 = a_0 I_0^2 + a_1 I_0 I_1 + a_2 I_0 I_2.
\end{equation}
What can be said in favor of the polynomial search is that if the zero mode is unstable, then it will register a crossing. If $I_{\mathcal{F}_+} I_{\mathcal{F}_-} < 0$, then one of the two factors must have the opposite sign of $I_0$. (Since $\mathcal{F}_+ (\mu = 1) = 0$, a nuance here involves whether the polynomials must be non-negative or strictly positive.) But suppose that the factors $I_{\mathcal{F}_+}$ and $I_{\mathcal{F}_-}$ are both negative. Then, although the zero mode is stable, the search registers a crossing anyway. Testing each factor individually is not equivalent to evaluating their product.

Thus the polynomial method neither generalizes nor even reproduces the zero-mode criterion (or for that matter the criteria in Eqs.~\eqref{pend1} and \eqref{pend2}). The difference between the two is in fact a crucial distinction. Eq.~\eqref{zero_mode} implies that the discriminant of the quadratic dispersion relation is negative, hence that $\textrm{Im}~\Omega \neq 0$ for a particular mode. The criterion $I_\mathcal{F} I_0 < 0$ lacks any such interpretation. 

To summarize, the following instability tests are found in the literature:
\begin{enumerate}
\item Linear stability analysis for all (or a range of) $\Omega$ and/or $K$.
\item $G(\mu_1) G(\mu_2) < 0$ for $\mu_1, \mu_2 \in [-1, 1]$: the point-by-point method of searching for angular crossings.
\item Eq.~\eqref{IFI0}: the polynomial method of searching for angular crossings.
\item Eq.~\eqref{zero_mode}: the zero-mode test (\textit{i.e.}, linear stability analysis for $K = K_0$).
\item Eq.~\eqref{pend1}: the $K = 0$ resonant-trajectory test.
\item Eq.~\eqref{pend2}: the $K = 0$ unstable-pendulum test.
\end{enumerate}
The presence of an angular crossing has been shown to be necessary and sufficient for the existence of an instability $\textrm{Im}~ \Omega \neq 0$ for some wave vector \cite{morinaga2021}. Besides that equivalence, no two of these tests produce identical results in all situations.

The point-by-point and polynomial methods are similar in that they adopt angular crossings as the instability criterion, differing in how they search $G(\mu)$ for crossings. But it is a separate question, which we now turn to, how $G(\mu)$ is constructed in the first place.

\section{Closures \& completions \label{sec:closures}}

We are drawing a distinction in this paper between closures (which supply a finite number of angular moments) and completions (which supply an infinite number) of the neutrino distribution functions. The distinction is helpful here because it reflects the differing needs of transport calculations and fast-instability searches. 

Much attention has been devoted to closure in the context of radiative transfer (\textit{e.g.}, Refs.~\cite{murchikova2017, richers2020} and the studies they cite). Its importance can be seen immediately by taking angular moments of the Boltzmann equation (here assuming flat spacetime with spherical symmetry and a static background):
\begin{align}
\frac{\partial}{\partial t} E + \frac{1}{r^2} \frac{\partial}{\partial r} \left( r^2 F \right) = \dots, \notag \\
\frac{\partial}{\partial t} F + \frac{1}{r^2} \frac{\partial}{\partial r} \left( r^2 P \right) = \dots,
\end{align}
and so on ad infinitum. (As before, $E$ and $F$ are the energy density and flux density. $P$ is the radiative pressure.) Particle streaming alone couples the moments in an infinite hierarchy of equations.

With a savvy implementation of closure, an accurate solution is obtained for the transport of energy and momentum. Many different closure methods have been proposed, implemented, and analyzed. Some of them---variable Eddington tensor methods---solve the Boltzmann equation under certain simplifications to all orders in the moment expansion and could be used to output approximate angular distributions. We are interested, however, in simulations that in practice only output the lowest moments, regardless of how they are obtained. These include simulations that adopt algebraic Eddington tensor methods.

In such cases, accuracy in $E$ and $F$ does not guarantee accuracy in the reconstruction of angular distributions using solely those moments. In other words, while it may not be necessary to track higher moments for the purposes of approximately solving the Boltzmann equation, higher moments must sometimes be supplied if the angular distributions are to be approximated well.

\begin{figure}
\centering
\includegraphics[width=.43\textwidth]{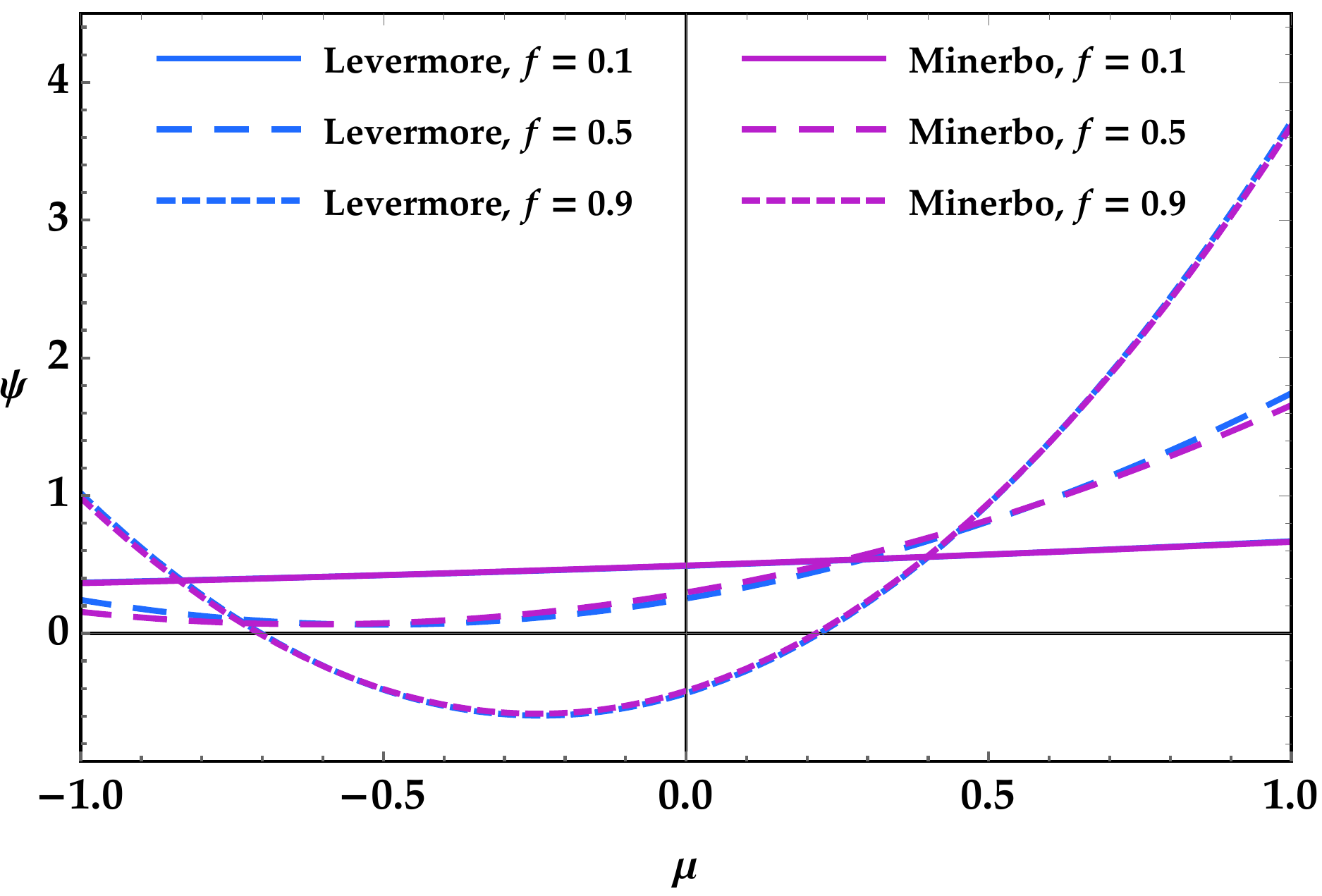}
\caption{Angular distributions obtained by applying M1 closure for the radiative pressure and setting all higher moments to zero. Distributions are normalized such that $\int d\mu ~\psi (\mu) = 1$.}
\label{levermore_minerbo}
\end{figure}

This point is illustrated in Fig.~\ref{levermore_minerbo}, which shows the angular distributions $\psi$ that result from applying two different M1 closure prescriptions. Three different flux factors $f$ are shown for each closure. In one spatial dimension, the radiative pressure is $P = p E$, where $p$ is the Eddington factor. Shown in the figure are the Levermore closure \cite{levermore1984}
\begin{equation}
p = \frac{3 + 4 f^2}{5 + 2 \sqrt{4 - 3 f^2}}~~[\textrm{Levermore}]
\end{equation}
and the Minerbo closure \cite{minerbo1978}
\begin{equation}
p = \frac{1}{3} + \frac{2 f^2}{15} \left( 3 - f + 3 f^2 \right)~~[\textrm{Minerbo}].
\end{equation}
These prescriptions follow from distinct physical assumptions. In the Levermore case, closure is obtained by asserting that the pressure is isotropic in the frame in which the flux vanishes. In the Minerbo case, the pressure is calculated by assuming that the underlying distribution function is classical and entropy-maximizing.

From Fig.~\ref{levermore_minerbo} we see that the distribution functions do not radically differ for the plotted flux factors. In particular, they both exhibit the same unphysical features at $f \gtrsim 0.5$: negative particle densities in non-radial directions and positive but highly inflated particle densities in the backward direction. 

Accurate modeling of forward-peaked distributions demands that higher moments be specified.

\section{Maximum-entropy \& sharp-decoupling distributions \label{sec:completions}}

To illustrate the latitude in modeling angular distributions---even with the densities and fluxes taken as given---we examine two completions in this section. For simplicity we consider monochromatic neutrinos, so that the angular distributions of number density and energy density have the same shape.

The first is the maximum-entropy distribution $\psi^{\textrm(max)}$, which maximizes the entropy $s = \psi \log \psi$ at fixed $E$ and $F$ \cite{minerbo1978, fort1997}. It has the form
\begin{equation}
\psi^{\textrm{(max)}} (\mu) = \frac{1}{e^{\eta - a \mu} + k},
\end{equation}
where $k = 0$. The distribution takes on $k=+1$ (Fermi--Dirac statistics) or $k = -1$ (Bose--Einstein) if the entropy functional is modified accordingly, but here we focus strictly on the classical limit as a representative case. For a distribution of this form, the flux factor is calculated to be
\begin{equation}
f^{\textrm{(max)}} = \coth a - \frac{1}{a},
\end{equation}
which can be numerically inverted to find $a(f)$. The other parameter $\eta$ sets the overall normalization of $\psi^\textrm{(max)}$. Normalizing so that the 0th moment is unity, we have
\begin{equation}
e^{-\eta} = \frac{a}{2 \sinh a}. \label{maxnorm}
\end{equation}
The maximum-entropy assumption is the origin of the Minerbo closure and its quantum-statistical variations \cite{janka1992, cernohorsky1994, smit2000}. Although $\psi^{\textrm{(max)}}$ models astrophysical neutrino angular distributions imperfectly, it does capture some of the essential features.

\begin{figure}{
\centering
\includegraphics[width=.43\textwidth]{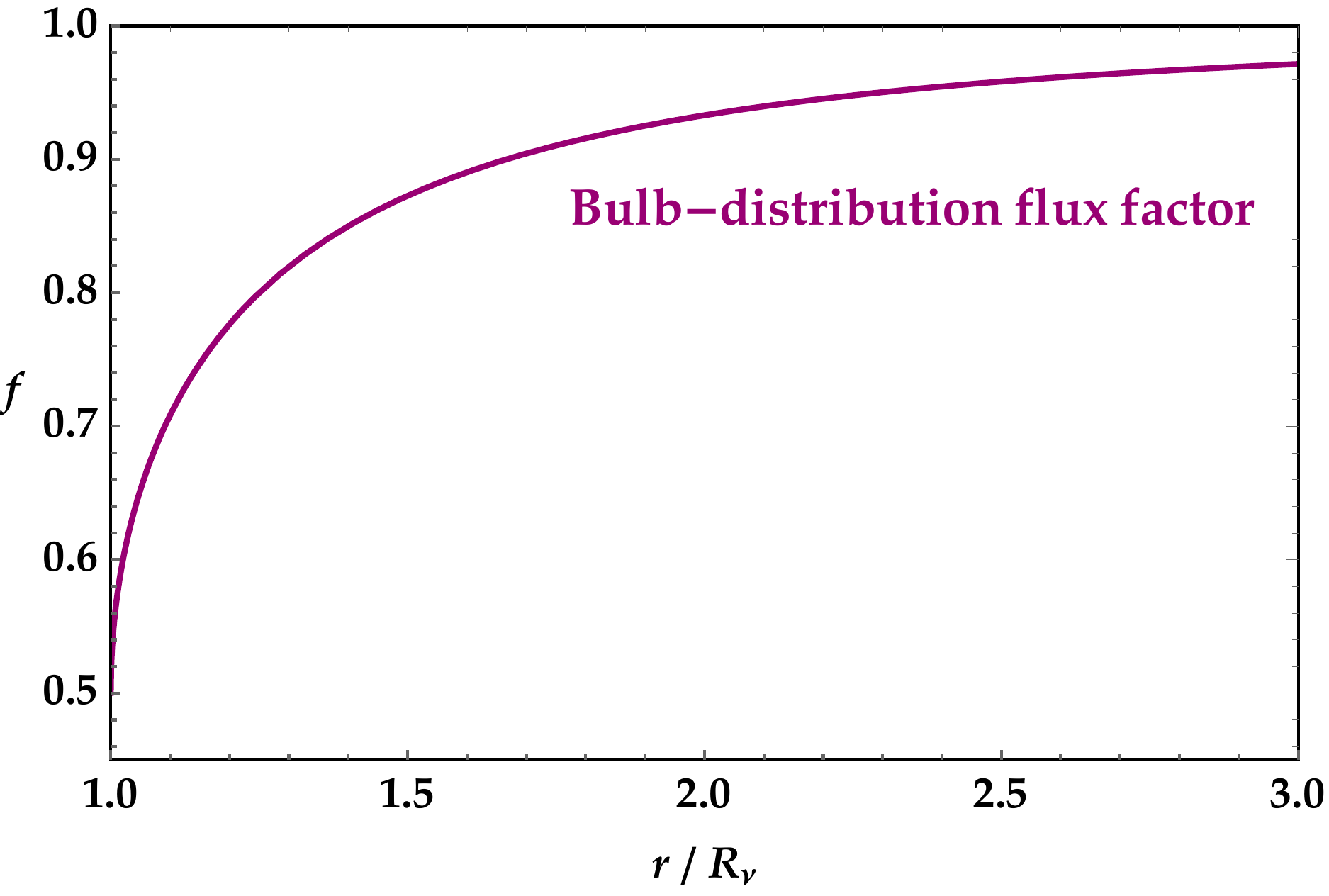}
}
\caption{Flux factor $f$ of the bulb distribution as a function of radius $r$ in units of the neutrinosphere radius $R_\nu$. Note that $f$ is never below $0.5$. Compare to Fig.~6 of Ref.~\cite{richers2017}, which shows the flux factor vs. radius in simulations with Boltzmann neutrino transport.}
\label{bulb_f_vs_r}
\end{figure}

An alternative way to specify all higher moments is to use a sharp-decoupling completion. A distribution of this type is found in the bulb model of supernova neutrino emission, in which all neutrinos decouple at a single sharp surface (the neutrinosphere) at radius $R_\nu$ \cite{duan2010}. The distribution is
\begin{equation}
\psi^\textrm{(bulb)} (\mu) = \mathcal{A} \theta \left( \mu - \sqrt{1 - \left( R_\nu / r \right)^2} \right),
\end{equation}
where $r$ is the radial coordinate, $\theta$ is the step function, and $\mathcal{A}$ is a normalization factor. We choose
\begin{equation}
\mathcal{A} = \frac{1}{1 - \sqrt{1 - 4 ( f - f^2)}}, \label{bulbnorm}
\end{equation}
which ensures as before that the 0th moment is unity. The flux factor in the bulb model is in one-to-one correspondence with the radial distance $r$:
\begin{equation}
f^\textrm{(bulb)} = \frac{\frac{1}{2} \left( \frac{R_\nu}{r} \right)^2}{1 - \sqrt{1-\left( \frac{R_\nu}{r} \right)^2}}
\end{equation}
or, inverting,
\begin{equation}
\frac{r}{R_\nu} = \frac{1}{2 \sqrt{f - f^2}}.
\end{equation}
The flux factor $f^\textrm{(bulb)}$ is plotted in Fig.~\ref{bulb_f_vs_r} as a function of $r / R_\nu$. It starts at the neutrinosphere with a value of $0.5$ and by twice the emission radius already exceeds $0.9$. Because neutrinos are radiated strictly outward, $\psi^{\textrm{(bulb)}}$ is intrinsically forward-peaked.

\begin{figure*}
\begin{subfigure}{
\centering
\includegraphics[width=.43\textwidth]{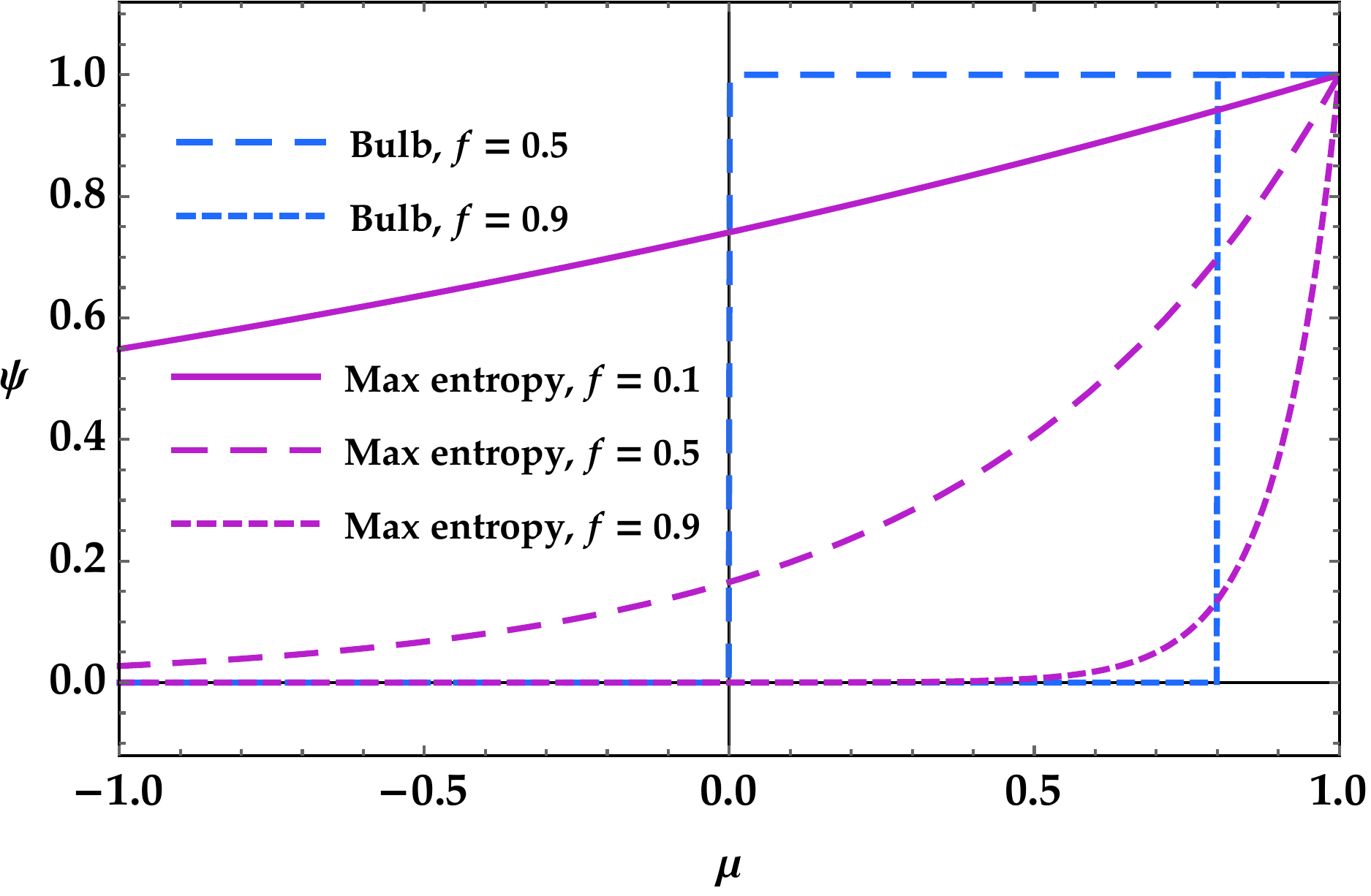}
}
\end{subfigure}
\begin{subfigure}{
\centering
\includegraphics[width=.43\textwidth]{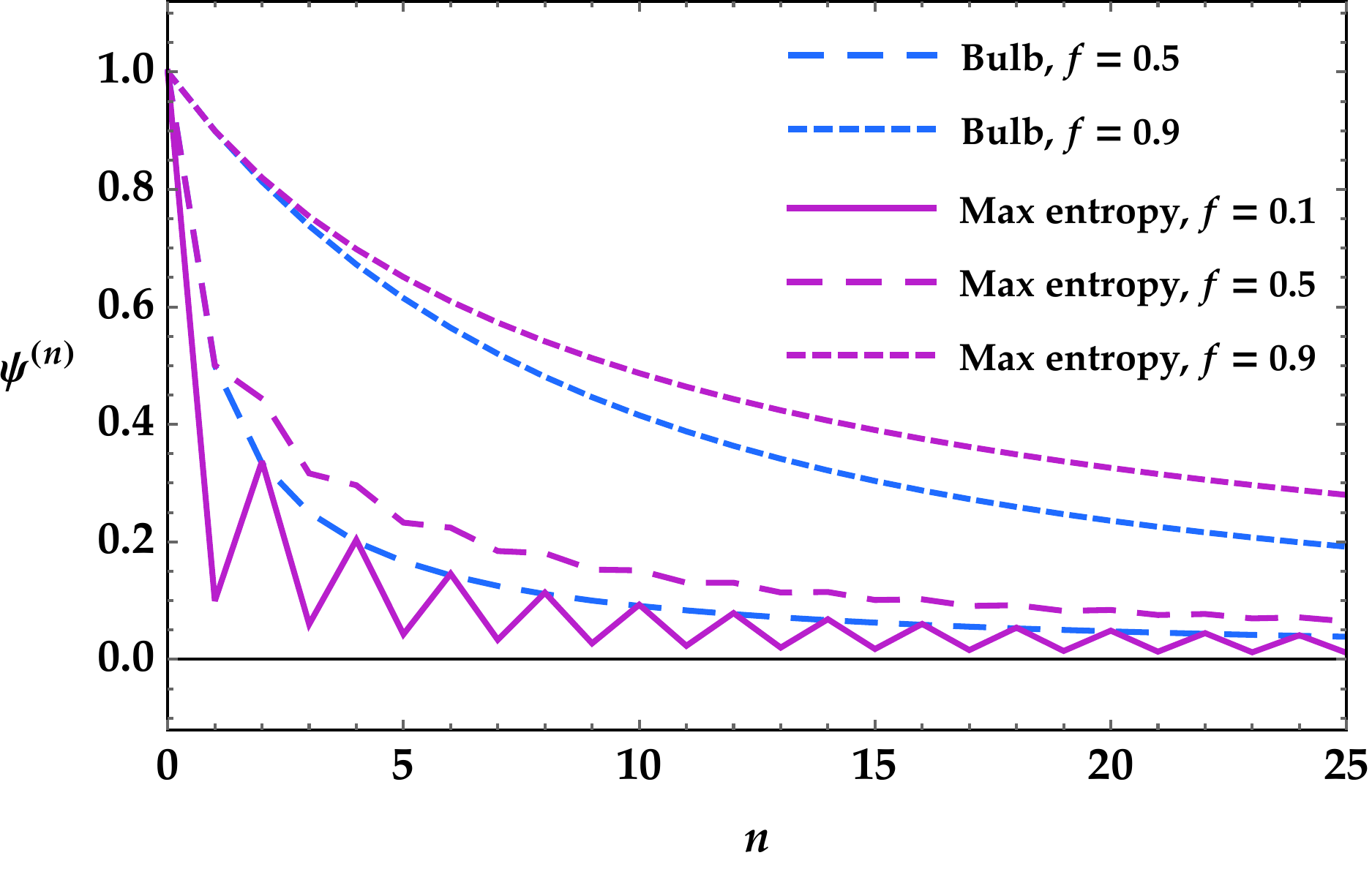}
}
\end{subfigure}
\caption{\textit{Left:} Angular distributions $\psi$ obtained using the sharp-decoupling (``Bulb'') and maximum-entropy completions. Unlike in the other figures, here the distributions are normalized to have $\psi (\mu = +1) = 1$. Note that $f = 0.5$ and $f = 0.9$ translate to picking bulb distributions at $r = R_\nu$ and $r = 1.67 R_\nu$, respectively. No bulb distribution exists with $f = 0.1$. \textit{Right:} Angular moments $\psi_n$, normalized to have $\psi_0 = 1$.}
\label{bulb_plots}
\end{figure*}

\begin{figure*}
\begin{subfigure}{
\centering
\includegraphics[width=.43\textwidth]{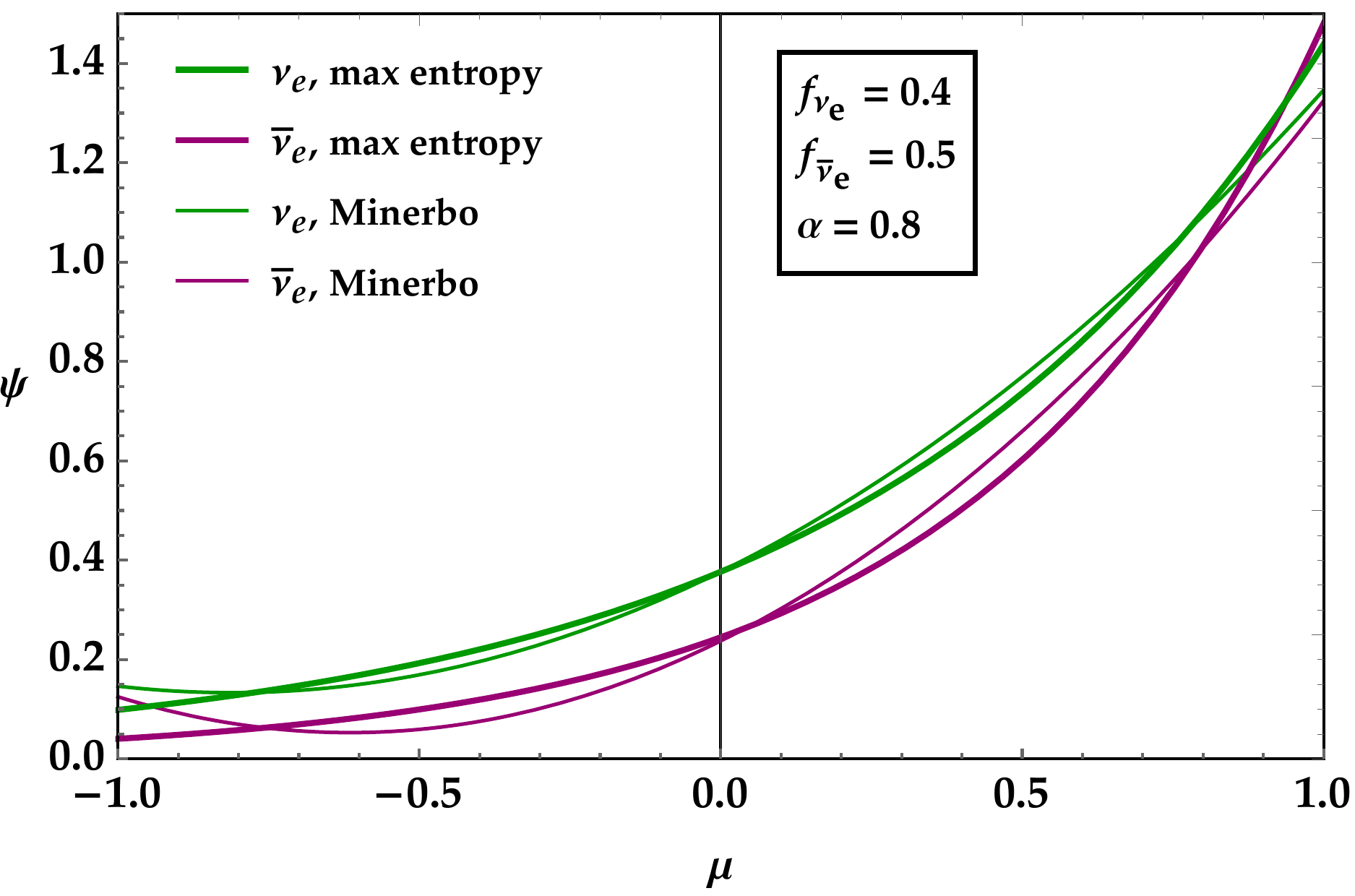}
}
\end{subfigure}
\begin{subfigure}{
\centering
\includegraphics[width=.43\textwidth]{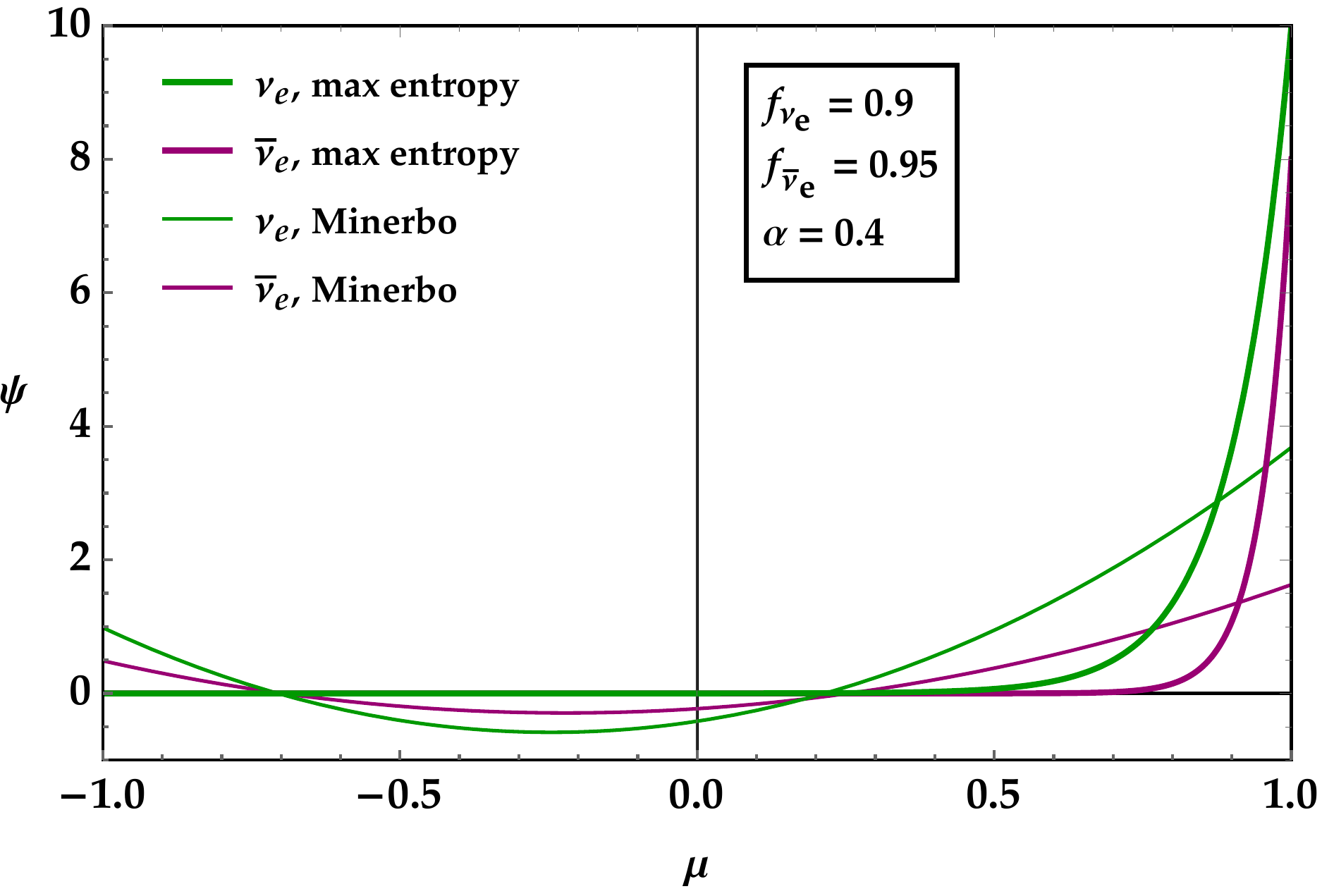}
}
\end{subfigure}
\caption{Angular distributions $\psi$ obtained using maximum-entropy completion and Minerbo closure, normalized such that $\int d\mu ~\psi_{\nu_e} (\mu) = 1$ and $\int d\mu ~\psi_{\bar{\nu}_e} (\mu) = \alpha$. Here $f_{\nu_e}$ and $f_{\bar{\nu}_e}$ are the flux factors of $\nu_e$ and $\bar{\nu}_e$. The hierarchies between the two are chosen to be fairly large in order to more clearly illustrate the possible discrepancies between completions and closures. \textit{Left:} The completed distributions (thick curves) cross near $\mu \sim 1$, a crossing not exhibited by the closed distributions (thin curves). \textit{Right:} The closed distributions exhibit a spurious double crossing due to unphysical modeling of non-forward fluxes. The completed distributions never cross.}
\label{closure_missed_spurious}
\end{figure*}

These completions are both suitable for certain applications, but they differ markedly. The left panel of Fig.~\ref{bulb_plots} compares $\psi^\textrm{(max)}$ and $\psi^\textrm{(bulb)}$ at the same flux factors as shown in Fig.~\ref{levermore_minerbo}. (No bulb distribution exists with $f = 0.1$.) We thus see two very different completions of the angular distribution even at fixed $f$.

Another way to inspect the angular information contained in these completions is by observing their angular moments
\begin{equation}
\psi_n = \int_{-1}^1 d\mu ~\mu^n \psi (\mu).
\end{equation}
These are shown in the right panel of Fig.~\ref{bulb_plots}. In order to write a recursion formula for $\psi^{\textrm{(max)}}$, we note that
\begin{equation}
\int_{-a}^a dx ~x^n e^x = a^n \left( e^a + (-1)^n e^{-a} \right) - n \int_{-a}^{a} dx ~x^{n-1} e^x.
\end{equation}
This then leads to
\begin{equation}
\psi_n^{\textrm{(max)}} = \frac{1}{a} \left[ e^{-\eta} \left( e^a + (-1)^n e^{-a} \right) - n \psi_{n-1}^{\textrm{(max)}} \right].
\end{equation}
Note that with the normalization of Eq.~\eqref{maxnorm}, $\psi_n^{\textrm{(max)}} \rightarrow 1$ as $a \rightarrow \infty$ (corresponding to $f \rightarrow 1$) at finite $n$. In the other limit, as $a \rightarrow 0$, $\psi_n^{\textrm{(max)}}$ alternates between 0 (at odd $n$) and finite but declining values (at even $n$). This is because all even $n$ have some contribution from the isotropic moment, whereas all odd $n$ vanish by symmetry. If angular distributions \textit{are} to be expressed using truncation, it is better to truncate an expansion in orthogonal functions (\textit{e.g.}, Legendre polynomials) than an expansion in $\mu^n$ moments. Closure prescriptions make the distinction moot, but only in part, by ensuring agreement in the isotropic limit.

For the sharp-decoupling completion, using the normalization of Eq.~\eqref{bulbnorm}, we have
\begin{equation}
\psi^{\textrm{(bulb)}}_n = \frac{1}{n+1} \frac{1 - \left( 1 - 4 \left( f - f^2 \right) \right)^{\frac{n+1}{2}}}{1 - \sqrt{ 1 - 4 \left( f - f^2 \right)}}.
\end{equation}
As the figure makes clear, forward-peaked distributions converge slowly in $n$ and vary in exactly how they do so.

In Sec.~\ref{sec:closures} we saw that standard M1 closures become unphysical at flux factor $f \gtrsim 0.6$. By comparing the Minerbo closure and the max-entropy completion, Fig.~\ref{closure_missed_spurious} shows that the failure of closure to accurately describe forward-peaked angular distributions leads to closures being unreliable when used in the search for angular crossings.

The figure shows potential errors of two kinds. In the left panel, an angular crossing in the completed distributions is missing from the closed ones at $\mu$ just above $0.9$. The reason for this crossing being absent is, again, that closures struggle with forward-peaking. In order for closure to place a lot of weight at $\mu \sim 1$, it must impose unrealistic negative fluxes in more transverse directions, as we saw above. The right panel of Fig.~\ref{closure_missed_spurious} shows that this undesired consequence can introduce spurious crossings. Even in cases where closure and completion both have crossings, a single crossing in the completion is typically turned into a double crossing in the closure.

While $\psi^{\textrm{(max)}}$ is not necessarily an optimal approximation, it does capture the main feature---concentrated flux at $\mu \sim 1$---that underlies the inadequacies of closure. We conclude that an angular-crossing search using closure is susceptible to these errors.

\section{Discussion \label{sec:discussion}}

Several recent publications have searched for fast instabilities, using angular crossings as a proxy. This manner of searching faces technical challenges associated with the limited angular resolution of simulations. The root of the problem is simple: as angular distributions become more forward-peaked, more angular moments are required for their accurate approximation.

Closure---which, in practice, usually means that the angular-moment expansion is truncated at the pressure or heat-flux tensor---is inadequate for forward-peaked distributions. Completions---where the available moments are matched to a physically motivated angular spectrum---are a superior alternative, but the question arises exactly \textit{which} physics should be used as the motivation. As their names indicate, the sharp-decoupling and maximum-entropy distributions are based on assumptions regarding decoupling and entropy. They are easy to formulate and work with, but they are not necessarily the best choices. 

Elsewhere we will present phenomenological fits to the neutrino angular distributions in a supernova simulation using Boltzmann transport \cite{nagakura2021}. From these fits, a completion can be constructed that better reflects the actual distributions found in supernovae.

\begin{acknowledgements}
We wish to thank Sajad Abbar, whose work on searching for angular crossings inspired us to write this paper. We also acknowledge helpful and enjoyable conversations with Sam Flynn, Nicole Ford, Evan Grohs, Jim Kneller, Gail McLaughlin, Taiki Morinaga, Sherwood Richers, and Don Willcox. L.J. acknowledges support provided by NASA through the NASA Hubble Fellowship grant number HST-HF2-51461.001-A awarded by the Space Telescope Science Institute, which is operated by the Association of Universities for Research in Astronomy, Incorporated, under NASA contract NAS5-26555. H.N. acknowledges support from the U.S. Department of Energy Office of Science and the Office of Advanced Scientific Computing Research via the Scientific Discovery through Advanced Computing (Sci- DAC4) program and Grant DE-SC0018297 (subaward 00009650).
\end{acknowledgements}

\bibliography{all_papers}

\end{document}